\documentclass{emulateapj}
\usepackage{color} 
\usepackage[colorlinks,urlcolor=blue,citecolor=blue,linkcolor=blue]{hyper ref}

\usepackage{epsfig}
\usepackage{subfigure}
\usepackage{graphicx}
\usepackage{amsmath}
\usepackage{natbib}

\newcommand{\mb}{\boldsymbol}

\shorttitle{Grain-Modified Magnetic Diffusivities in PPDs}
\shortauthors{Xu \& Bai}

\begin{document}


\title{On the Grain-Modified Magnetic Diffusivity in Protoplanetary Disks}


\title{On the Grain-Modified Magnetic Diffusivities in Protoplanetary Disks}
\author{Rui Xu\altaffilmark{1,2}, Xue-Ning Bai\altaffilmark{3}}


\affil{$^1$ Department of Astrophysical Sciences, Princeton University, Princetion, NJ 08544;
ruix@princeton.edu}
\affil{$^2$ Yuanpei College, Peking University, Beijing, 100871, China}
\affil{$^3$ Institute for Theory and Computation, Harvard-Smithsonian Center for Astrophysics, 
60 Garden St., MS-51, Cambridge, MA, 02138; xbai@cfa.harvard.edu}




\begin{abstract}
Weakly ionized protoplanetary disks (PPDs) are subject to non-ideal-magnetohydrodynamic (MHD)
effects including Ohmic resistivity, the Hall effect and ambipolar diffusion (AD), and the resulting
magnetic diffusivities ($\eta_O, \eta_H$ and $\eta_A$) largely control the disk gas dynamics.
The presence of grains not only strongly reduces disk ionization fraction, but also modify the scalings
of $\eta_H$ and $\eta_A$ with magnetic field strength. We derive analytically asymptotic expressions
of $\eta_H$ and $\eta_A$ in both strong and weak field limits and show that towards strong field,
$\eta_H$ can change sign (at a threshold field strength $B_{\rm th}$), mimicking a flip of field polarity,
and AD is substantially reduced.
Applying to PPDs, we find that when small $\sim0.1$ ($0.01$)$\mu$m grains are sufficiently abundant
[mass ratio $\sim0.01$ ($10^{-4}$)], $\eta_H$ can change sign up to $\sim2-3$ scale heights above
midplane at modest field strength (plasma $\beta\sim100$) over a wide range of disk radii. Reduction
of AD is also substantial towards the AD dominated outer disk and may activate the
magneto-rotational instability. We further perform local non-ideal MHD simulations of the inner disk
(within 10 AU) and show that with sufficiently abundant small grains, magnetic field amplification due
to the Hall-shear instability saturates at very low level near the threshold field strength $B_{\rm th}$.
Together with previous studies, we conclude by discussing the grain-abundance-dependent
phenomenology of PPD gas dynamics.
\end{abstract}



\section{Introduction}

Gas dynamics of the weakly ionized protoplanetary disks (PPDs) is to a large extent controlled
by its coupling with magnetic fields, described by non-ideal magnetohydrodynamic (MHD)
effects, including Ohmic resistivity, the Hall effect and ambipolar diffusion (AD). The three effects
dominate in different regions of PPDs, and affect the gas dynamics in different ways. A constant
theme of PPD research has been to estimate the strength of magnetic diffusivities, and study
their consequences to gas dynamics, especially on the level of turbulence and efficiency of
angular momentum transport (see \citealp{Turner_etal14} for recent review).

The strength of magnetic diffusivities largely depends on gas density $\rho$, level of ionization,
and magnetic field strength $B$. In the absence of grains, the Ohmic, Hall and ambipolar
diffusivities can be expressed in particularly simple form \citep{SalmeronWardle03}
\begin{equation}\label{eq:diff1}
\eta_O=\eta_e\ , \quad\eta_H=\frac{cB}{4\pi en_e}\ ,\quad
\eta_A=\frac{B^2}{4\pi\gamma_i\rho\rho_i}\ ,
\end{equation}
where $\eta_e\propto (n/n_e)$ is resistivity due to electrons, $n_e$, $n$ are the electron and
neutral number densities, and $\rho_i$ is the ion density.

These expressions are complicated by the presence of dust grains. On the one hand, grains
can substantially reduce the level of ionization, and hence boosting all three magnetic
diffusivities proportionally. On the other hand, grains carry charge and result in much more
complex dependence of diffusivities on magnetic field strength (e.g., see \citealp{Wardle07}
and \citealp{Bai14e}), yet such complex dependence has not been systematically explored.

As an initial effort, \citet{Bai11b} (hereafter B11) considered the limit when grains are tiny
($\sim$nm sized PAHs) so that the conductivity of charged grains is similar to ions.
He showed that when tiny grains are the dominant charge carrier, Hall diffusivity diminishes
and AD is substantially reduced in the strong field limit, which may be important in the outer
region of PPDs.

In this paper, we extend the work of B11 to more general situations with more typical grain
sizes (sub-micron to micron) relevant to PPDs. We focus on the disk interior (rather than
surface) where grains are likely the dominant charge carrier, and are mostly singly charged,
and systematically explore how magnetic diffusivities depend on field strength.
We hope to provide researchers a useful guidance on modeling non-ideal MHD
effects from ionization chemistry, which has been fruitful in the recent years
(e.g., \citealp{Turner_etal07,BaiStone13b,Bai13,Bai14,Bai15,Lesur_etal14,Gressel_etal15,Simon_etal15b}).

This work is also motivated by the recent advances in understanding the gas dynamics of
the Hall effect, where the Hall-shear instability may substantially amplify the midplane field
strength in the inner region of PPDs \citep{Kunz08,Lesur_etal14,Bai14} when the background
field is aligned with disk rotation. With strong field amplification, we discuss whether
grain-modified diffusivities can feedback to PPD gas dynamics.

We present our simplified model in Section \ref{sec:model}, with application to PPDs
discussed in Section \ref{sec:app}. We perform MHD simulations of the inner disk and
discuss the results in Section \ref{sec:sim}. In Section \ref{sec:sum}, we summarize
and conclude.

\section{A Simplified Model for Grain-modified Diffusivities}\label{sec:model}

In weakly ionized gas, the general expressions for the three magnetic diffusivities are
\citep{Wardle07,Bai11a}
\begin{equation}\label{eq:diffu}
\begin{split}
&\eta_O=\frac{c^2}{4\pi}\frac{1}{\sigma_O}\ ,\quad
\eta_H=\frac{c^2}{4\pi}\frac{\sigma_H}{\sigma_H^2+\sigma_P^2}\ ,\\
& \eta_A=\frac{c^2}{4\pi}\frac{\sigma_P}{\sigma_H^2+\sigma_P^2}-\eta_O\ ,
\end{split}
\end{equation}
where $\sigma_O, \sigma_H, \sigma_P$ are Ohmic, Hall and Pederson conductivities,
defined as
\begin{equation}
\begin{split}\label{eq:sigfull}
& \sigma_O=\frac{ec}{B}\sum_j n_j|Z_j|\beta_j\ ,\\
& \sigma_H=\frac{ec}{B}\sum_j\frac{n_j Z_j}{1+\beta_j^2}\ ,\\
& \sigma_P=\frac{ec}{B}\sum_j\frac{n_j |Z_j|\beta_j}{1+\beta_j^2}\ ,
\end{split}
\end{equation}
where summation goes over all charged species $j$, with $n_j$, $Z_je$ being the
number density and charge of individual charged species. The Hall parameter
$\beta_j$ is given by
\begin{equation}\label{eq:hallpara}
\beta_j=\frac{|Z_j|eB}{m_j c}\frac{1}{\gamma_j\rho}\ ,
\end{equation}
which describes the ratio between the gyro-frequency of the charged species and
its collision frequency with the neutrals, where
$\gamma_j\equiv\langle\sigma v\rangle_j/(\mu_n + m_j)$, with $\langle\sigma v\rangle_j$
being rate coefficient for collisional momentum transfer with the neutrals. The charged
species $j$ is strongly coupled to the neutrals if $\beta_j\ll 1$ (weak field), and it is strongly
coupled to magnetic fields when $\beta_j\gg 1$ (strong field). The values of
$\langle\sigma v\rangle_j$ for electron, ion and grain can be found in Equation (14)-(16) of
\citep{Bai11a,Bai14e}, giving Hall parameter
\begin{equation}\label{eq:beta}
\begin{split}
\beta_e & \approx2.1(B/n_{15})\min\ [1,\ T_{100}^{-1/2}]\ ,\\
\beta_i & \approx  3.3\times10^{-3}(B/n_{15})\ ,\\
\beta_g & \approx (B/n_{15})\min\ [3.2\times10^{-3}|Z_g|^{-1},\ 
10^{-9}a_1^{-2}T_{100}^{-1/2}]
\end{split}
\end{equation}
where $B$ is magnetic field measured in Gauss, $n_{15} = n_H/10^{15} \rm{cm^3}$,
$Z_g$ is grain charge, $a_1$ is grain size in $\mu m$, disk temperature
$T_{100}=T//100{\rm K}$. Clearly, $\beta_e\gg\beta_i$, and unless grain size is tiny
($a\lesssim$nm), $\beta_i\gg\beta_g$.

As long as ion mass is much larger than neutral mass, different ion species have their Hall
parameters very close to $\beta_i$ quoted above, and can be treated as a single species.
In the interior of PPDs where ionization level is low, chemistry calculations show that (small)
grains are largely singly charged until the ionization fraction exceeds grain abundance
towards disk surface (e.g., Figure 6 of \citealp{Wardle07} and Figure 1 of \citealp{Bai11a}).
We also assume single-sized grains for simplicity, which is typically adopted in PPD
chemistry calculations. While a grain size distribution is expected in reality, we expect a
single-size treatment to capture and better clarify the essence of their effects.
Therefore, we consider only four types of charged species: electrons, ions, and
positive/negative charged grains, whose number densities are represented by
$n_e$, $n_i$, $n_{\rm gr}^{+}$ and $n_{\rm gr}^-$.

Under these conditions, the Hall and Pederson conductivities read
\begin{equation}\label{eq:con}
\begin{split}
\sigma_H= &\frac{-1}{1+\beta_e^2}+\frac{n_i}{n_e}\frac{1}{1+\beta_i^2}+\frac{n_{gr}^{+}-n_{gr}^{-}}{n_e}\frac{1}{1+\beta_g^2}\ ,\\
\sigma_P = &\frac{\beta_e}{1+\beta_e^2}+\frac{n_i}{n_e}\frac{\beta_i}{1+\beta_i^2}
+\frac{n_{\rm gr}^{+}+n_{\rm gr}^{-}}{n_e}\frac{\beta_g}{1+\beta_g^2}\ .
\end{split}
\end{equation}
Hereafter, we omit the prefactor $en_ec/B$ in conductivities and $cB/4\pi en_e$ in diffusivities
so as to make them dimensionless. The grain-free diffusivities (\ref{eq:diff1}) corresponds
to $\eta_O=1/\beta_e$, $\eta_H=1$ and $\eta_A=\beta_i$, and hence the conventional
Ohmic regime applies in weak field with $\beta_e<1$, AD regime applies in strong field
$\beta_i>1$, and the Hall dominated regime lies in between. For brevity, we further define
$n_{\rm gr}^\pm\equiv n_{\rm gr}^++n_{\rm gr}^-$.

Charge neutrality gurantees that only three of the four number densities are independent.
With additional normalization by $n_e$, only two of them serve as independent model
parameters. We find it useful to choose the two ratios $n_i/n_e$,
$n_{\rm gr}^\pm/n_i$ as model variables. Without grains, their values are
obviously $1$ and $0$, while when grains become important charge carriers,
$n_i\gg n_e$, and $n_{\rm gr}^\pm>n_i$.

The Hall parameters $\beta_{e,i,g}\propto B$ can be considered as a proxy for magnetic
field strength. According to (\ref{eq:beta}), we fix $\beta_i=10^{-3}\beta_e$ in our calculations.
The ratio $\beta_g/\beta_i\propto a^{-2}$ reflects grain size and defines our last (third) model
parameter. For $0.1\mu$m sized grains, $\beta_g\sim10^{-4}\beta_i$, and only for
$a\sim10^{-3}\mu$m we have $\beta_i\sim\beta_g$.

\begin{figure*}[!ht]
\centering
\includegraphics[width =170mm]{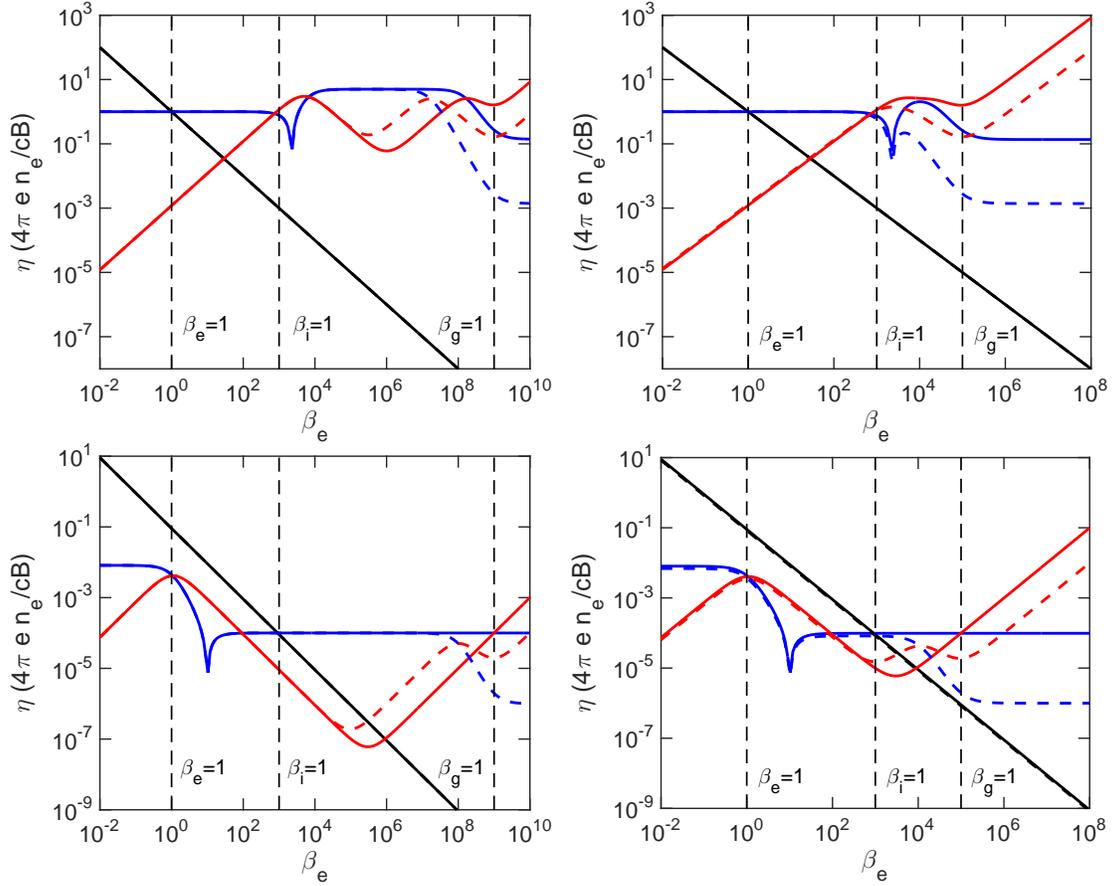}
\caption{Dimensionless Ohmic (black), Hall (blue), and the ambipolar (red) diffusivities as a
function of magnetic field strength (manifested as the electron Hall parameter $\beta_e$) under
various representative parameters. We set $n_i/n_e=1.2$ and $10^4$ in the top and bottom panels,
respectively. On the left two panels, we have $\beta_g /\beta_i = 10^{-6}$ ($\mu$m sized grain),
while in right two panels we have $\beta_g/\beta_i=10^{-2}$ ($\sim0.01\mu$m sized grain).
In each panel, solid lines correspond to $(n_{gr}^+ + n_{gr}^-)/n_i = 1$ ,while the dashed lines
correspond to $(n_{gr}^+ + n_{gr}^-)/n_i = 10$.}
\label{fig:analytic_diff}
\end{figure*}

\subsection[]{Asymptotic Relations}

We are mainly interested in reducing (\ref{eq:con}) to sufficiently simple and intuitive
expressions. Ohmic resistivity is the most straightforward and can be written as
\begin{equation}
\eta_O=\frac{1}{\beta_e(1+\theta_1+\theta_2)}\ ,
\end{equation}
where we have defined
\begin{equation}
\theta_1 = \frac{n_i}{n_e}\frac{\beta_i}{\beta_e}\ ,\quad
\theta_2 = \frac{n_{\rm gr}^\pm}{n_e}\frac{\beta_g}{\beta_e}\ .
\end{equation}
In general, Ohmic resistivity results from electron conductivity. Contributions from the ions
and grains become significant ($\theta_{1,2}\gtrsim1$) only when
$n_i/n_e>\beta_e/\beta_i\approx1000$ and
$n_{\rm gr}^{\pm}/n_e\gtrsim\beta_e/\beta_g$, both requiring grain charge density to be
overwhelmingly dominant.

Calculations of the Hall and ambipolar diffusivities involve much more complex algebra,
and sufficiently simple expressions can only be obtained in asymptotic regimes.
In the weak field regime (equivalently, Ohmic dominated regime) where
$\beta_g\lesssim\beta_i\ll\beta_e\ll 1$, we Taylor expand the expressions for $\sigma_H$,
$\sigma_P$ in equation (\ref{eq:con}) and keep the leading order terms\footnote{We
have also assumed $n_i/n_e\ll10^6$, or $\beta_i\theta_1\ll\beta_e$, which is almost
always the case from chemistry calculations.}. The results are
\begin{equation}\label{eq:redu_con}
\sigma_H \approx \beta_e^2\ , \quad
\sigma_P \approx \beta_e[(1+\theta_1+\theta_2)-\beta_e^2]\ .\\
\end{equation}
Substituting them to Equation (\ref{eq:diffu}), we obtain
\begin{equation}
\begin{split}
\eta_H&\approx\frac{1}{(1+\theta_1+\theta_2)^2}\ .\\
\eta_A & \approx \frac{\beta_e (\theta_1+\theta_2)}{(1+\theta_1+\theta_2)^3}
=\beta_i\frac{n_i+n_{\rm gr}^\pm(\beta_g/\beta_i)}{n_e(1+\theta_1+\theta_2)^3}\ .
\end{split}
\end{equation}
These expressions reduces exactly to Equation (15) of B11 ($\beta_g=\beta_i$)
by replacing $\theta_1+\theta_2$ by his $\theta$, and replacing
$n_i+n_{\rm gr}^\pm(\beta_g/\beta_i)$ by his $\bar{n}\equiv n_i+n_{\rm gr}^{\pm}$.
As already discussed there, ambipolar diffusivity is enhanced by the presence of
charged grains, and the Hall diffusivity can be modestly reduced when
$\theta_{1,2}\gtrsim1$.


In the strong field limit where  $1\ll\beta_g\lesssim\beta_i\ll\beta_e$, we can
reduce the Hall and Pederson conductivities to
\begin{equation}
\begin{split}
\sigma_H &\approx\frac{\theta_1\beta_e}{\beta_i}
\bigg[\frac{1}{\beta_i^2}-\frac{1}{\beta_g^2}\bigg]+\frac{1}{\beta_g^2}\ ,\\
\sigma_P &\approx\beta_e\bigg[\frac{\theta_1}{\beta_i^2}+\frac{\theta_2}{\beta_g^2}\bigg]\ .
\end{split}
\end{equation}
The corresponding Hall and ambipolar diffusivities are
\begin{equation}\label{eq:eta_a}
\begin{split}
\eta_H&\approx\frac{\left[\theta_1\beta_e(\beta_g^2-\beta_i^2)+\beta_i^3\right]\beta_i\beta_g^2}{\beta_e^2(\beta_g^2\theta_1+\beta_i^2\theta_2)^2}\ ,\\
\eta_A&\approx\frac{\beta_i^2\beta_g^2}{\beta_e(\theta_1\beta_g^2+\theta_2\beta_i^2)}\ .
\end{split}
\end{equation}
In the tiny grain limit of $\beta_g = \beta_i$, these relatively complex expressions can be further
reduced to
\begin{equation}
\begin{split}
\eta_H &\approx\frac{\beta_i^2}{\beta_e^2\theta^2}=\left(\frac{n_e}{\overline{n}}\right)^2\ ,\\
\eta_A &\approx\frac{(\theta_1+\theta_2)\beta_i^6}{\beta_e\beta_i^4(\theta_1+\theta_2)^2}=\frac{\beta_i n_e}{\overline{n}}\ ,
\end{split}
\end{equation}
which agree exactly with equation (16) of B11.

From Equation (\ref{eq:eta_a}), we see that in the strong field regime, $\eta_H$ becomes
negative when
\begin{equation}\label{eq:sign}
\theta_1\beta_e(\beta_g^2-\beta_i^2)+\beta_i^3<0\ ,
\ {\rm or}\ \ 
\beta_g^2<\beta_i^2(1-\frac{n_e}{n_i})\ .
\end{equation}
In general, the sign of $\eta_H$ depends on the relative mass (mobility) between the
positive and negative charge carrier. In the absence of grains, with electrons much more
mobile than ions, $\eta_H$ is positive. In the presence of tiny grains with
$\beta_g=\beta_i$ as considered by B11, charged grains are as mobile as normal ions.
Therefore, negative charge carriers are still effectively more mobile due to the presence
of electrons, and hence $\eta_H\geq0$. Relation (\ref{eq:sign}) states that, as long as
grains are not too small (i.e. well above nanometer size so that $\beta_g\ll\beta_i$),
$\eta_H$ can become negative whenever there are more ions than electrons. In other
words, there are more negatively charged grains than positively ones, and hence
negative charge carriers are effectively less mobile. We will see in Section
\ref{sec:app} that $n_i>n_e$ almost always holds from ionization chemistry
calculations in PPDs. The main practical question then is whether disk magnetic
field can reach the level such that $\eta_H$ changes sign. We therefore define
$B_{\rm th}$ as the threshold field strength beyond which $\eta_H$ changes sign,
and this is a main quantity of interest that we study in the rest of the paper.

\subsection[]{Representative Cases}

In Figure \ref{fig:analytic_diff}, we show the dimensionless Ohmic, Hall and the ambipolar
diffusivities as a function of magnetic field strength (characterized by the electron Hall
parameter), covering most of the parameter space relevant for grain size
$a\gtrsim0.01\mu$m (see figure caption).
We first confirm that $\eta_H\propto B$ and $\eta_A\propto B^2$ holds both in weak
($\beta_e\ll1$) and strong ($\beta_g\gg1$) field regimes as expected. In between, we
see that both $\eta_H$ and $\eta_A$ exhibit complex variations with $B$, where
$\eta_H$ changes sign in all cases, and $\eta_A$ is reduced in steps compared with the
$B^2$ scaling.

Top two panels of Figure \ref{fig:analytic_diff} correspond to cases where grains
are modestly important charge carrier, giving $n_i/n_e$ not far from order unity.
We see that for field strength with $\beta_i<1$, $\eta_H$ and $\eta_A$ behave
normally as in conventional Ohmic and Hall regimes. Sign change of $\eta_H$
and reduction of $\eta_A$ are both achieved towards stronger field (in the
conventional AD regime) with $\beta_i>1$ (and $\beta_g<1$).

On the other hand, in the bottom two panels, assuming grains are the dominant
charge carriers such that $n_i/n_e\gg1$, we see that magnetic diffusivities
behaves qualitatively differently. First, the threshold field strength $B_{\rm th}$
for $\eta_H$ to change sign is much weaker, close to $\beta_e=1$. Similarly,
reduction of $\eta_A$ starts right from $\beta_e=1$. Secondly, Ohmic resistivity
dominates all the way through the conventional Hall-regime ($\beta_e>1$,
$\beta_i<1$), while the Hall effect dominates only in much stronger field
($\beta_i>1$ and $\beta_g<1$), which is in the conventional AD-dominated
regime. AD dominates only when the field is much stronger with
$\beta_g>1$.

Overall, we see that the ratio $n_i/n_e$ largely controls the dependence of
$\eta_H$ and $\eta_A$ on magnetic field strength, especially towards weaker
field ($\beta_i\lesssim1$). Other parameters (grain size and
$n_{\rm gr}^\pm/n_i$) mainly affect diffusivity behaviors towards strong field
with $\beta_i>1$, leading to complex dependence before reaching the
asymptotic regime (\ref{eq:eta_a}) at $\beta_g>1$, and trend can be identified
by comparing the left and right panels of Figure \ref{fig:analytic_diff}.

\begin{figure*}[!ht]
\centering
\includegraphics[width =170mm]{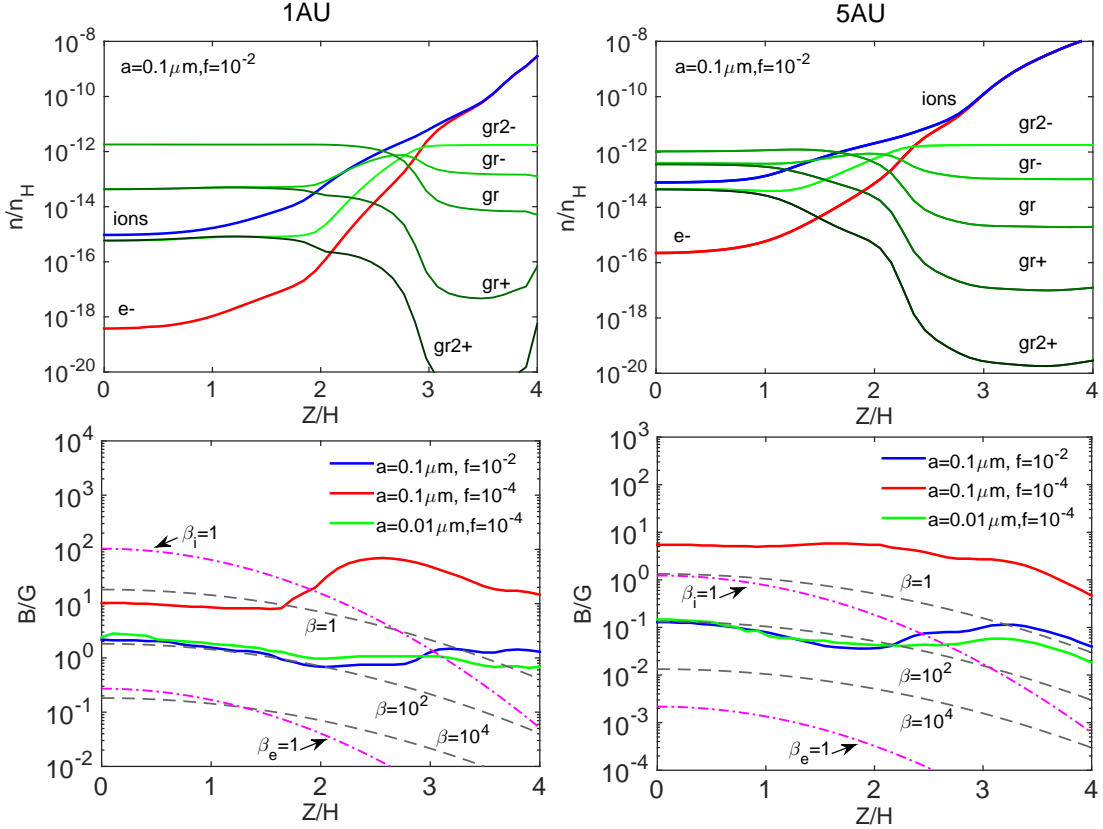}
\caption{Results from ionization chemistry calculations at 1 AU (left) and 5 AU (right).
Top panels: abundance of all charged species as a function of height $z$
above disk midplane (normalized to disk scale height $H$), where gr$n$(+/-) denote
positively/negatively charged grains with grain charge $n$, and the red/blue lines
represent abundances of electrons/all ions.
Bottom panels: solid lines show threshold field strength $B_{\rm th}$ (beyond which
$\eta_H$ changes sign, in Gauss, see legend)
as a function of $z$ based on the chemistry calculations in the top panels.
Also shown are the field strengths that correspond to plasma $\beta=1$,
100 and $10^4$ (grey), as well as field strengths that give electron/ion Hall parameters
$\beta_e=1$ and $\beta_i=1$ (pink). Three combinations of grain size $a$ and dust-to-gas
mass ratios $f$ are considered in the bottom panels, as marked in legend, while the top
panels correspond to the $a=0.1\mu$m and $f=10^{-2}$ case.}
\label{fig:abundance}
\end{figure*}

\section[]{Application to Protoplanetary Disks: Ionization Chemistry}\label{sec:app}

To test our analytical results, we conduct ionization-recombination chemistry calculation
using a complex reaction network developed in \citet{BaiGoodman09} and \citet{Bai11a}
that followed from \citet{IlgnerNelson06}. There are 175 gas-phase species and over two
thousands gas phase reactions extracted from the latest version of the UMIST database
\citep{UMIST12}. The rate coefficients in the data base are given as a function of
temperature. One change from our previous chemistry calculations is that when gas
temperature lies out of the stated range of validity, we still use the given formula as if
it remained valid (whereas originally we computed the coefficients using upper or lower
bound of valid temperature range). For inner disk temperatures ($\sim100-300$K),
about $10\%$ of the reactions are affected, majority of which are neutral-neutral reactions,
and/or reactions with large activation energy. In the latter case, our original approach
tends to significantly overestimate the reaction rates at lower temperatures. Also, for many
of these reactions, we note that the rate coefficients listed in the KIDA chemical
database generally have different valid temperature ranges with similar coefficients.
This change does not affect the level of ionization in the presence of grains (but leads to
higher ionization fraction in the grain-free case), while the composition of ion species is
affected.

We include a single grain population with maximum grain charge of $\pm2$, with fiducial
choices of grain size $a=0.1\mu$m and dust-to-gas mass ratio of $f=10^{-2}$. Although
such grain size and abundance are likely exaggerated considering that substantial grain
growth must have occurred in PPDs (e.g., \citealp{Birnstiel_etal10}), they are the best to
illustrate the effects that we discuss in this paper. Given that large uncertainties remain in
our understandings of grain size distribution in PPDs (e.g., \citealp{DAlessio_etal06}),
we also consider a few other combinations of grain sizes and abundances to study
parameter depdendnce. Note that for chemical purposes, we are only concerned with the
abundance of sub-micron sized grains, which is the most relevant to ionization and
charging. We adopt the minimum-mass solar nebular (MMSN) disk model that is vertically
isothermal with standard cosmic-ray and X-ray ionization prescriptions as
described in \citet{Bai11a}, together with a constant ionization rate of
$7\times10^{-19}$ s$^{-1}$ as a proxy for radioactive decay (the exact value matters little).

The relative importance of non-ideal MHD effects in PPDs is conveniently described by
the dimensionless Elsasser numbers, defined as
\begin{equation}
\Lambda\equiv\frac{v_A^2}{\eta_O\Omega_K}\ ,\quad
\chi\equiv\frac{v_A^2}{\eta_H\Omega_K}\ ,\quad
Am\equiv\frac{v_A^2}{\eta_A\Omega_K}\ ,
\end{equation}
where $v_A^2=B^2/4\pi\rho$ is the Alfv\'en speed, $\Omega_K$ is disk angular frequency.
Non-ideal MHD effects become dominant when the (absolute value of) respective Elsasser
numbers becomes smaller than order unity.
Note that in our definition, $\chi$ will change sign when $\eta_H$ becomes negative.

\begin{figure*}
\centering
\includegraphics[width =180mm]{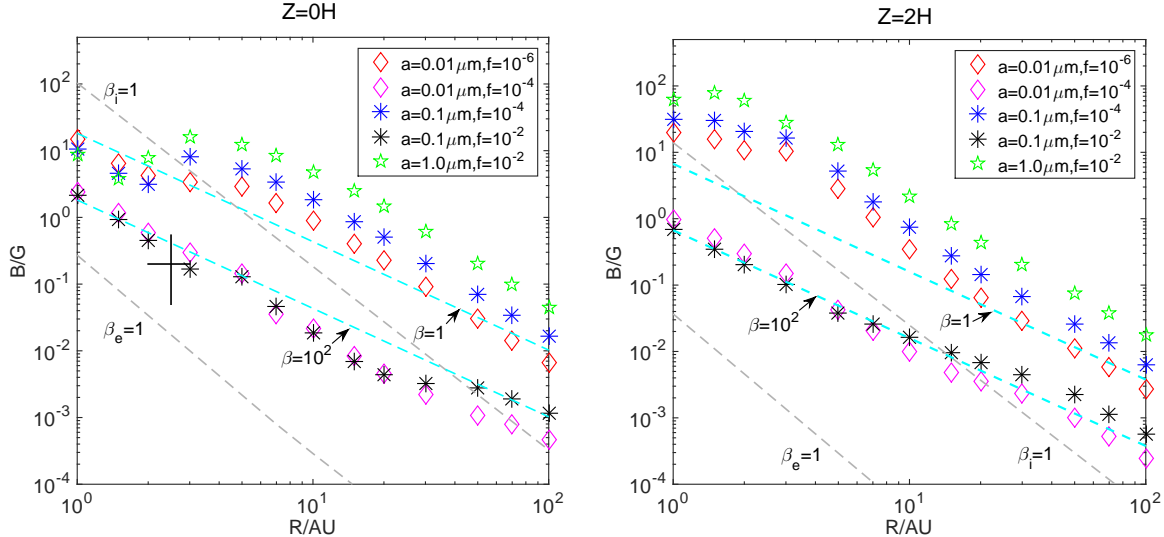}
\caption{Threshold field strength $B_{\rm th}$ (in Gauss) beyond which the Hall coefficient
changes sign, plotted as a function of disk radius $R$ from ionization chemistry
calculations at disk midplane (left) and two scale heights above midplane (right).
Five combinations of grain sizes $a$ and dust-to-gas mass ratios $f$ are shown,
as marked in the legend. To guide the eye, we also show field strengths that
correspond to plasma $\beta=1$ and 100 (black solid), as well as field strengths
that give electron/ion Hall parameters $\beta_e=1$ and $\beta_i=1$ (grey dashed).}
\label{fig:etaHsign}
\end{figure*}


\subsection[]{Threshold Field Strength in PPDs}

In Figure \ref{fig:abundance}, we show the vertical profile of charged species abundance
at 1 AU and 5 AU using fiducial grain size ($a=0.1\mu$m) and mass ratio ($f=10^{-2}$) in
the top two panels. We see that grains are the dominant charge carriers within $\sim2-3$
disk scale heights ($H$) about disk midplane, leading to large ratios of $n_i/n_e$. This
situation is similar to the solid curves in the two bottom panels of Figure \ref{fig:analytic_diff},
and we have confirmed that our simple four-species model well captures the $\eta_H$ and
$\eta_A$ behaviors computed from the full chemical abundances using (\ref{eq:sigfull}).
as long as most grains are singly charged.

In the bottom panels of Figure \ref{fig:abundance}, we further show the threshold field
strength $B_{\rm th}$ (beyond which $\eta_H$ changes sign) as a function of height above
midplane. 
To guide eye, we further plot field strengths that correspond to $\beta_e=1$ and $\beta_i=1$
(pink), as well as field strengths that give plasma $\beta$, ratio of gas pressure to magnetic
pressure, to be $1$, $100$ and $10^4$.

We first focus on the fiducial case with $a=0.1\mu$m and $f=0.01$ (blue lines). We see
that within $\sim2H$, $B_{\rm th}$ is well below the $\beta_i=1$ line,
consistent with our expectations given $n_i\gg n_e$. Towards the surface, as ionization
fraction approaches and then exceeds grain abundance, we see while $\eta_H$ can still
change sign, but at field strength well beyond $\beta_i=1$, analogous to the top panels of
Figure \ref{fig:analytic_diff}. We also note that the threshold field strength $B_{\rm th}$
corresponds to only modest level of magnetization, with $\beta\sim100$ within $\pm2H$
about midplane, which is very likely accessible in real PPDs (see next subsection).

Using different grain size and abundances yield different threshold field strengths
$B_{\rm th}$. Reducing dust-to-gas mass ratio by a factor of 100 to $f=10^{-4}$
increases $B_{\rm th}$ by about an order of magnitude, and we see that at both
1 AU and 5 AU, field strength near or above equipartition ($\beta=1$) is needed to
have $\eta_H$ change sign. This is much less likely to be achieved in PPDs, at
least under standard disk models where field strength well below equipartition
is sufficient to drive disk accretion at expected accretion rate (e.g., \citealp{Bai13}).
On the other hand, for the same dust-to-gas mass ratio $f=10^{-4}$, reducing the
grain size further to $0.01\mu$m yields results similar to the fiducial case. This is
because smaller grains have higher total surface area/abundance at given
mass, both are favorable to make charged grains dominant charge carriers.

\subsection[]{An Empirical Criterion on Grain Size and Abundance}

Extending the discussion to a broad range of disk radii, we show in Figure
\ref{fig:etaHsign} the threshold field strength $B_{\rm th}$ as a function of orbital
radius $R$, and present the results at disk midplane ($z=0$) and intermediate
height ($z=2H$). Moving towards larger disk radii, ionization fraction increases
systematically (due to reduced density and deeper penetration of external
ionization), as can be traced from Figure \ref{fig:abundance}. This makes
$B_{\rm th}$ shift up relative to the $\beta_i=1$ line as discussed before.
In the mean time, as gas density/pressure drops, field strength at fixed
plasma $\beta$ also shifts up relative to the $\beta_i=1$ line, as is
straightforward to show from their definitions ($B\propto\sqrt{\rho}$ v.s.
$B\propto\rho$). As a result, we see from Figure \ref{fig:etaHsign} that
line showing $B_{\rm th}$ as a function of $R$ is approximately parallel the
constant plasma $\beta$ line over a very wide range of radii. 

In Figure \ref{fig:etaHsign}, we have considered a broad collection of grain
sizes and abundances. We are particularly interested in the question of
for which combinations of grain sizes and abundances can $\eta_H$
change sign in the expected range of PPD magnetic field strength. We
consider plasma $\beta=100$ as a target field strength, which is easily
achieved in either pure magnetorotational instability (MRI, \citealp{BH91})
turbulence (e.g., \citealp{Davis_etal10}), or in more realistic studies of
PPD gas dynamics (e.g., \citealp{BaiStone13b,Lesur_etal14,Bai14,Bai15}).
On the other hand, field strength above equipartition is unlikely in the
disk interior. For reference, we also mark the possible field strength of the
solar nebular inferred from the Semekona meteorites \citep{Fu_etal14},
which likely corresponds to midplane field strength at $2-3$ AU during the
epoch of chondrule formation. Although large uncertainties remain, field
strength of up to $\beta=100$ in an MMSN disk is reasonably consistent
with the inferred nebular field strength.

From Figure \ref{fig:etaHsign}, it is clear that small grains in large abundance
is essential to make $B_{\rm th}$ sufficiently weak (plasma $\beta\gtrsim100$)
that can potentially be achieved in PPDs. In particular, the results from our
fiducial choice of $a=0.1\mu$m, $f=10^{-2}$ are very similar to those obtained
with the $a=0.01\mu$m, $f=10^{-4}$ case, at both $z=0$ and $z=2H$. This
might suggest an empirical criterion where $f/a^2$ should be at least order
unity (with $a$ measured in $\mu$m) in order to achieve a sufficiently weak
threshold field $B_{\rm th}$. Note that a constant $f/a^2$ corresponds to
a constant $na$ where $n$ is grain number density. Therefore, the controlling
parameter is not the total grain surface area ($\propto na^2$), but lies in
between total surface area and number density. We might further generalize
this consideration into a distribution of grain sizes, and define
\begin{equation}
G\equiv\int\frac{df/da}{(a/{\mu}{\rm m})^2}da\ .\label{eq:g}
\end{equation}
Our discussions above suggest $G\gtrsim1$ to achieve sufficiently weak
$B_{\rm th}$.

\begin{figure*}[!ht]
\centering
\includegraphics[width =180mm]{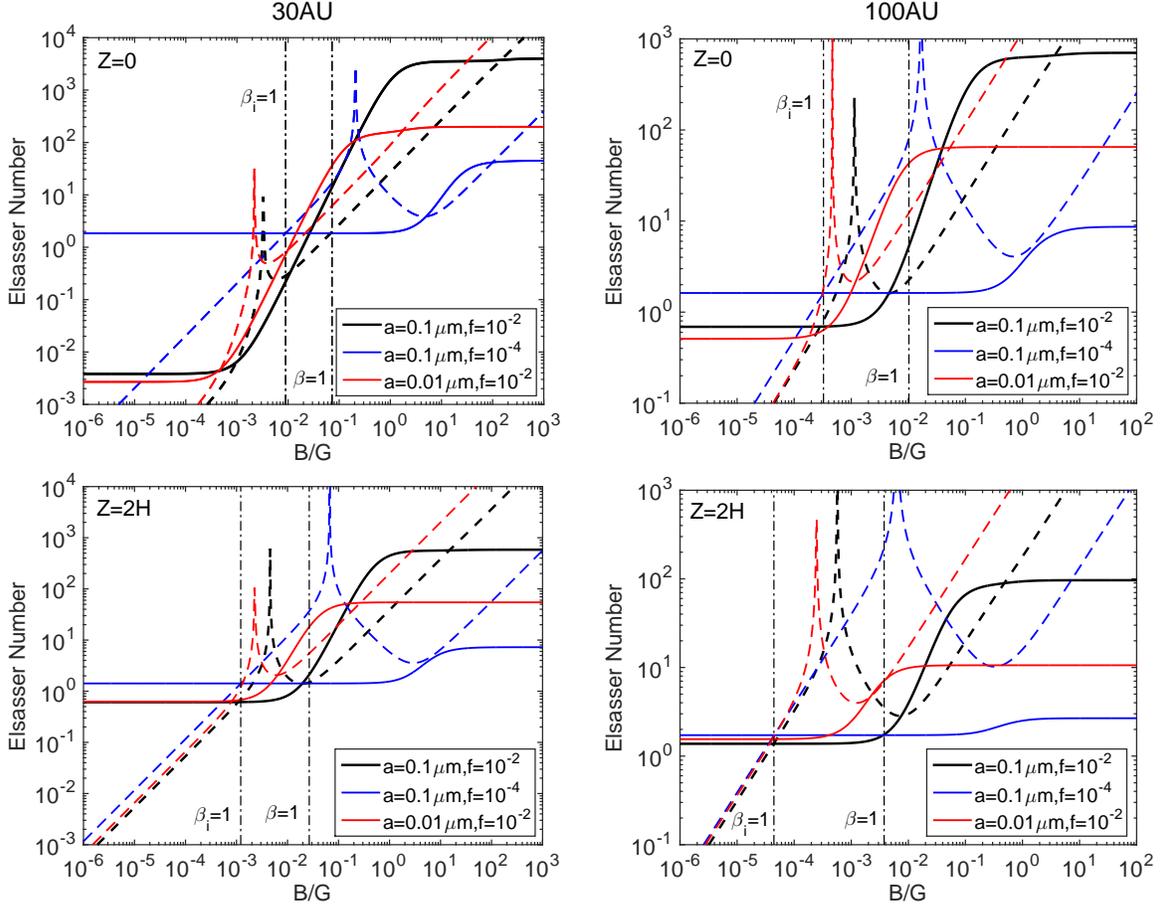}
\caption{The ambipolar Elsasser number $Am$ (solid) and Hall Elsasser number
$\chi$ (dashed) as a function of field strength (in Gauss) from ionization chemistry
calculations at 30 AU (left) and 100 AU (right). Top and bottom panels show the
results from calculations at disk midplane, and at two scale heights above midplane,
respectively. Three combinations of grain sizes $a$ and dust-to-gas mass ratios $f$
are shown, as marked in the legends.  We also mark the field strength corresponding
to plasma $\beta=1$, as well as field strengths that give ion Hall parameters
$\beta_i=1$, in vertical dash-dotted lines.}\label{fig:etaAreduce}
\end{figure*}

We note that when assuming the standard MRN size distribution for
interstellar grains \citep{MRN77}, with $dn/da\propto a^{-3.5}$ between
$a_{\rm min}\approx0.005\mu$m and $a_{\rm max}\approx0.25\mu$m,
we obtain $G\approx22\gg1$. Therefore, while grain growth in PPDs must
substantially reduces $G$ from interstellar value, we do expect that
grain-modified diffusivities is likely relevant at least in some stages of
PPD evolution.

\subsection[]{Diffusivities in the Outer Disk}

Although $\eta_H$ can change sign essentially at all distances, it is mainly
the inner disk ($\lesssim10$ or at most a few tens of 10 AU) that is the
most relevant. Conventionally, $\beta_i>1$ marks the dominance of AD
over the Hall effect, which is achieved at lower density regions towards the
outer disk. The complex dependence of $\eta_A$ and $\eta_H$ on field
strength in the presence of abundant small grains complicates the situation.
In Figure \ref{fig:etaAreduce}, we show the AD and Hall Elsasser numbers
$Am$ and $\chi$ as a function of magnetic field strength at 30 AU and 100 AU,
at both disk midplane and intermediate height ($z=2H$). We consider three
combination of grain sizes and abundances as before and discuss the results
below.

We first see that $Am$ is constant at both weak and strong field regimes, as
expected. In the presence of abundant small grains ($G\gtrsim1$, black and red
lines), the difference in $Am$ between the two regimes can reach several orders
of magnitude. In between at intermediate field strength, we have
$\eta_A\approx$constant or $Am\propto B^2$ (see also
Figure \ref{fig:analytic_diff}). Therefore, AD behaves the same way as Ohmic
resistivity (except for its anisotropy). These trends are the same as those
discussed in B11 in the presence of tiny ($a\sim$nm) grains.
If we expect field strengths in the outer PPDs correspond to plasma
$\beta\sim1-10^4$, then roughly speaking, the midplane region up to $\sim100$
AU, and upper layer ($z=2H$) up to $\sim30$ AU would lie in this intermediate
regime when $G\gtrsim1$.

With less amount of small grains ($G\ll1$, blue lines in Figure \ref{fig:etaAreduce}),
on the other hand, reduction of AD occurs at much stronger field. For both 30 AU and
100 AU, both at midplane and a few scale heights above, such reduction requires
field strength above equipartition level, thus is unlikely to be relevant. In fact, for
our choice of $a=0.1\mu$m and $f=10^{-4}$, the ionization fraction at disk midplane
already exceeds grain abundance, and hence the role of grains is only minor.
In this case, we see that $Am\sim1$ is a good proxy over a wide range of disk radii
and vertical locations, consistent with previous findings
\citep{Bai11a,Bai11b,PerezBeckerChiang11a}.

We also see that in the presence of abundant small grains, substantial reduction of
AD towards strong field ($\beta_i>1$) is accompanied by sign change of $\eta_H$,
and the Hall effect can dominate over AD (in the conventional AD dominated regime).
This is consistent features seen in Figure \ref{fig:analytic_diff} (e.g., bottom panels).
Therefore, the Hall effect may play a more important role than commonly assumed.

\begin{table*}
\begin{center}
\caption{List of all quasi-1D simulations.}\label{tab:runs}
\begin{tabular}{c|c|c|c|c|c|c|c|c}\hline\hline
 $R$ (AU) &  $\beta_0$ & $f$ & Diffusivity  & $\beta_{\rm mid}$  & $\alpha^{\rm Max}$ & $T_{z\phi}^{z_b}$ & $\dot{M}_w$   & $z_b$\\\hline
 
 1 & $10^5$ & $10^{-2}$ & full & $9.50\times 10^3 $ & $4.00\times10^{-4}$ & $1.76\times10^{-4}$ & $2.02\times10^{-5}$ &  3.94             \\
 
 1 & $10^5$ & $10^{-2}$ & simple & $8.54\times10^3$ & $5.69\times10^{-4}$ & $1.95\times10^{-4}$ & $3.15\times10^{-5}$ &     3.86       \\
 
 1 & $10^5$ & $10^{-4}$ & any & $8.10\times10^2$ & $5.96\times10^{-3}$ & $2.67\times10^{-4}$ & $6.27\times10^{-5}$   &     4.05            \\\hline
 
 1 & $10^4$ & $10^{-2}$ & full & $5.38\times10^3$ & $1.25\times10^{-3}$ & $7.46\times10^{-4}$ & $4.57\times10^{-5}$ &       3.83             \\
 
 1 & $10^4$ & $10^{-2}$ & simple & $4.62\times10^3$ &  $2.09\times10^{-3}$ & $7.84\times10^{-4}$ & $7.22\times10^{-5}$&      4.00         \\     
 
 1 & $10^4$ & $10^{-4}$ & any & $4.81\times10^2$ & $1.80\times10^{-2}$ & $1.24\times10^{-3}$ & $1.30\times10^{-4}$ & 3.95                  \\\hline
 
 5 & $10^5$ & $10^{-2}$ & full & $5.58\times10^3$ & $2.04\times10^{-4}$ & $2.70\times10^{-4}$ & $4.97\times10^{-5}$&      3.94             \\
 
 5 & $10^5$ & $10^{-2}$ & simple & $3.41\times10^3$ & $4.62\times10^{-4}$ & $3.32\times10^{-4}$ & $6.70\times10^{-5}$ &  3.91           \\
 
 5 & $10^5$ & $10^{-4}$ & any & $2.77\times10^2$ & $5.20\times10^{-3}$ & $4.99\times10^{-4}$ & $1.43\times10^{-4}$&    3.91         \\\hline
 
 5 & $10^4$ & $10^{-2}$ & full & $2.61\times10^3$ & $1.13\times10^{-3}$ & $1.22\times10^{-3}$ & $1.61\times10^{-4}$&       3.83            \\
 
 5 & $10^4$ & $10^{-2}$ & simple & $2.08\times10^3$ & $1.88\times10^{-3}$ & $1.30\times10^{-3}$ & $1.86\times10^{-4}$&      3.91            \\
 
 5 & $10^4$ & $10^{-4}$ & any & $4.61\times10^2$ & $1.75\times10^{-2}$ & $1.83\times10^{-3}$ & $2.29\times10^{-4}$&       3.71           \\\hline
 
 \hline
\end{tabular}
\end{center}
Note: we use grain size $a=0.1\mu$m for all runs, and ``any" means ``full" or ``simple"
diffusivity treatments yield same results. See Section \ref{ssec:setup} for more details.
\end{table*}


The AD Elsasser number $Am$ directly controls the level of the MRI turbulence,
and $Am\sim1$ is generally considered to be sufficiently strong to substantially
damp the MRI \citep{BaiStone11,Simon_etal13a,Simon_etal13b}. Further
introducing a modestly strong Hall term would enhance the level of turbulence if
net vertical field is aligned with disk rotation ${\mb B}_0\cdot{\mb\Omega}>0$,
while reduce it for opposite polarity
\citep{SanoStone02b,KunzLesur13,Bai15,Simon_etal15b}. The grain-modified
magnetic diffusivities in the outer disk studied here may lead to two interesting
effects when $G\gtrsim1$. First, the fact that AD behaves similarly as Ohmic
resistivity at intermediate field strength may lead to cycles of growth
and decay of the MRI as observed in resistive MRI simulations \citep{Simon_etal11}.
Second, the sign change of $\eta_H$ means that the enhancement of the MRI
by the Hall effect would occur at anti-aligned field polarity, rather than aligned polarity.
These effects deserve further investigation via three-dimensional MHD simulations.

\section{MHD simulations of the inner PPD}\label{sec:sim}

\subsection[]{Simulation Setup and Parameters}\label{ssec:setup}

To further study the effect of grain-modified magnetic diffusivities on PPD gas
dynamics, we follow \citet{Bai14} (hereafter B14) and conduct local stratified
shearing-box MHD simulations that include all three non-ideal MHD terms. Note
that B14 simply assumed $\eta_H \propto B$, $\eta_A \propto B^2$ in the
calculations, and used a 2D diffusivity table showing the proportional coefficients
as a function of density and ionization rate to compute the diffusivities. To
incorporate the complex dependence of $\eta_H$ and $\eta_A$ on magnetic field
strength, we extend the diffusivity table to 3D that explicitly accounts for this
dependence.

Here we mainly focus on the inner disk region ($R<10$ AU), where
it suffices to conduct quasi-1D simulations\footnote{It means that the
simulation box is largely 1D in vertical, but allows for $4$ cells in radial
and azimuthal dimensions, which facilitates the simulations to relax to
a steady state.} (as verified in \citealp{Bai15}) because the MRI is largely
suppressed. The simulations use isothermal equation of state, and the
simulation box extends from $-8H$ to $8H$ in $z$, with a resolution of 18
cells per $H$. We consider examples at two radii $R=1$ AU and $5$ AU,
using diffusivity tables with fixed grain size $a=0.1\mu$m but two different
abundances $f=10^{-2}$ and $10^{-4}$. An artificial boost of ionization rate
beyond column densities of $\sim0.03$ g cm$^{-2}$ is included to mimic
the effect of FUV ionization that brings the gas to ideal MHD regime. All
simulations are conducted
with net vertical magnetic field $B_{z0}$ threading the disk, and the field
strength is parameterized by plasma $\beta_0$ defined as ratio of
midplane gas pressure to magnetic pressure of the net vertical field.
We consider $\beta_0=10^4$ and $10^5$, as adopted in most previous
works chosen to yield accretion rate in the expected range for the
MMSN disk model. In code units, we have midplane density
$\rho_{\rm mid}=1$, isothermal sound speed $c_s=1$, Keplerian
frequency $\Omega_K=1$, and magnetic permeability $\mu_m=1$.

It is well known that the effect of the Hall term depends on the polarity
of background vertical field $B_{z0}$ threading the disk (e.g.,
\citealp{Wardle99,BalbusTerquem01,WardleSalmeron12}). Of particular
interest is the case where ${\mb B}_0\cdot{\mb\Omega}>0$, or background
field being aligned with rotation axis. Linear analysis as well as numerical
simulations have shown that in this case, horizontal components of the
magnetic field can be strongly amplified due to the Hall shear instability
(HSI, \citealp{Kunz08,Lesur_etal14,Bai14}). Because both radial and
azimuthal components are amplified, Maxwell stress ($-B_RB_\phi$)
is strongly enhanced which likely plays a non-negligible role in disk angular
momentum transport. On the other hand, when
${\mb B}_0\cdot{\mb\Omega}<0$, the opposite occurs, and the horizontal
field is reduced to close to zero near the midplane region \citep{Bai14,Bai15}.
Because the interesting effects we have discussed in this paper operate
when magnetic field gets amplified, we consider only aligned polarity
${\mb B}_0\cdot{\mb\Omega}>0$ in our simulations.

\begin{figure*}
\centering
\includegraphics[width =180mm]{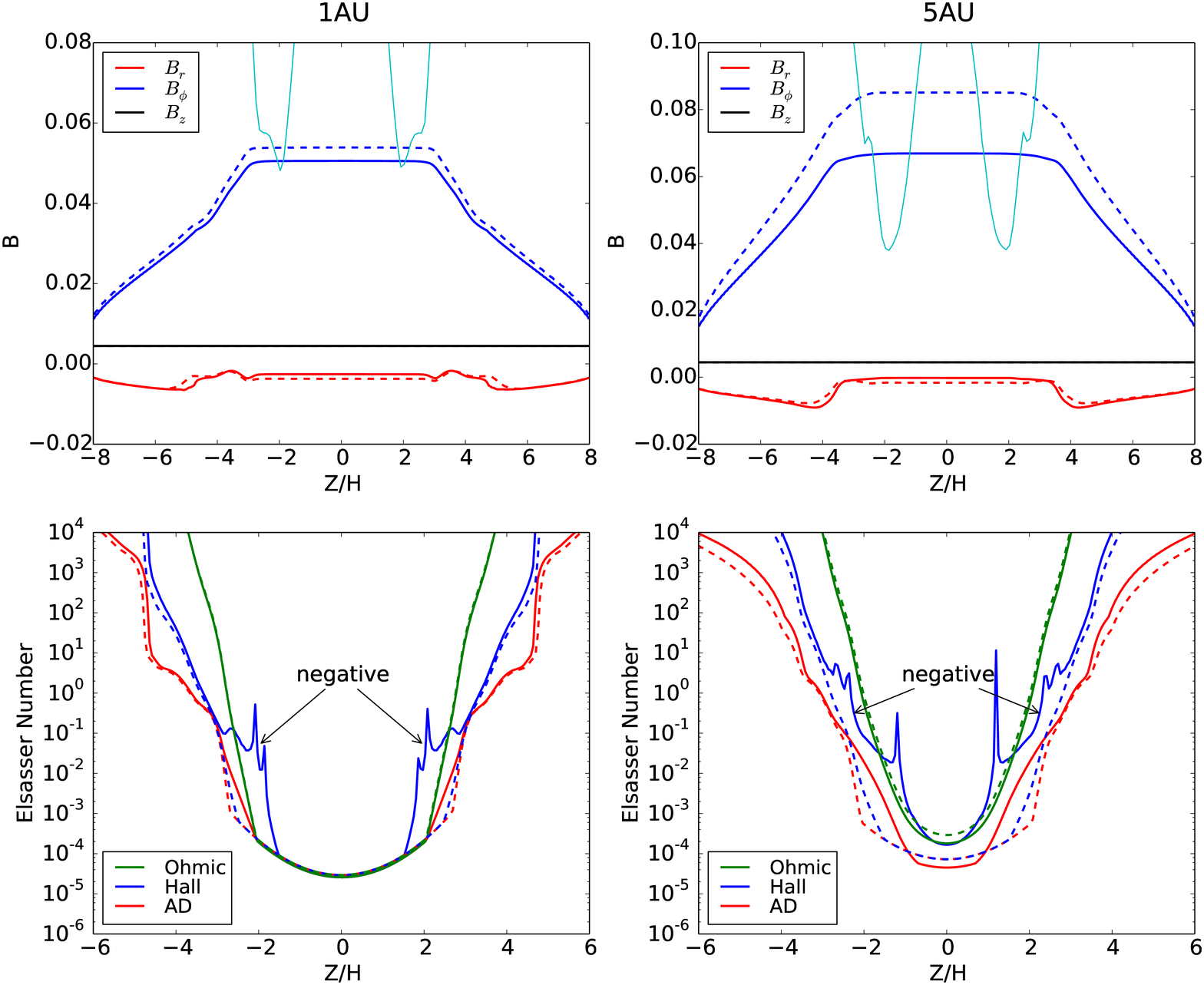}
\caption{Top: vertical profiles for three components of magnetic field strength
(in code units) from simulations at 1AU (left) and 5 AU (right) in steady state,
where equipartition field strength at midplane corresponds to $B=\sqrt{2}$.
We have chosen grain size $a=0.1\mu$m and dust-to-gas mass ratio $f=0.01$, with
weak net vertical field $\beta_0=10^5$.
Dashed lines correspond to simulations assuming $\eta_H\propto B$ and
$\eta_A \propto B^2$ in the weak field limit, while solid lines (except the cyan line)
correspond to simulations using full dependence of $\eta_H$ and $\eta_A$ on $B$.
Cyan solid lines mark the threshold field strength above which $\eta_H$ changes
sign. Bottom: vertical profiles of Ohmic, Hall and AD Elsasser numbers from
simulations using full dependence of $\eta_H$ and $\eta_A$ on $B$
at 1 AU (left) and 5 AU (right).}
\label{fig:simulation}
\end{figure*}

We run the simulations following the procedures of \citet{Bai14}. We first
turn off the Hall effect during initial evolution to time $t=120\Omega^{-1}$.
Then we turn on the Hall effect and run the simulations to
$t=240\Omega^{-1}$, where steady state magnetic configuration fully
saturates. In all cases, the saturated field geometry obeys ``odd-z"
symmetry (e.g., Figure 9 of \citealp{BaiStone13b}) where horizontal
components of magnetic field remain the same sign across the disk,
and reach maximum around disk midplane. While this symmetry is
unphysical for disk wind launching, it is likely an artifact of shearing-box
and here we set this issue aside\footnote{Whether a physical ``even-z"
symmetry can be obtained in shearing-box also depends on the diffusivity
profiles. With $a=0.1\mu$m and $f=10^{-4}$, \citet{Bai14,Bai15} obtained
a physical wind geometry at $R=5$ AU with $\beta_0=10^5$. We can
reproduce this result but we fail to obtain physical symmetry with $f=10^{-2}$.}.
Because magnetic diffusivities in the disk midplane region are always
excessively large at the inner disk, a diffusivity cap $\eta_{\rm cap}$ for
each non-ideal MHD term is implemented to prevent prohibitively small
numerical timestep. Most previous works used
$\eta_{\rm cap}\lesssim10H^2\Omega$. We use
$\eta_{\rm cap}=10H^2\Omega$ in the runs up to $t=240\Omega^{-1}$
as described before, but then enlarge $\eta_{\rm cap}$ to $100H^2\Omega$
and run for another $\Delta t=60\Omega^{-1}$, which appears sufficient for
the system to relax to a new state. We do observe changes in $B_R$ at
midplane by up to $30\%$ after enlarging the diffusivity cap, while typically
the changes are within $10\%$, and is much smaller for midplane $B_\phi$.

\begin{figure*}
\centering
\includegraphics[width =180mm]{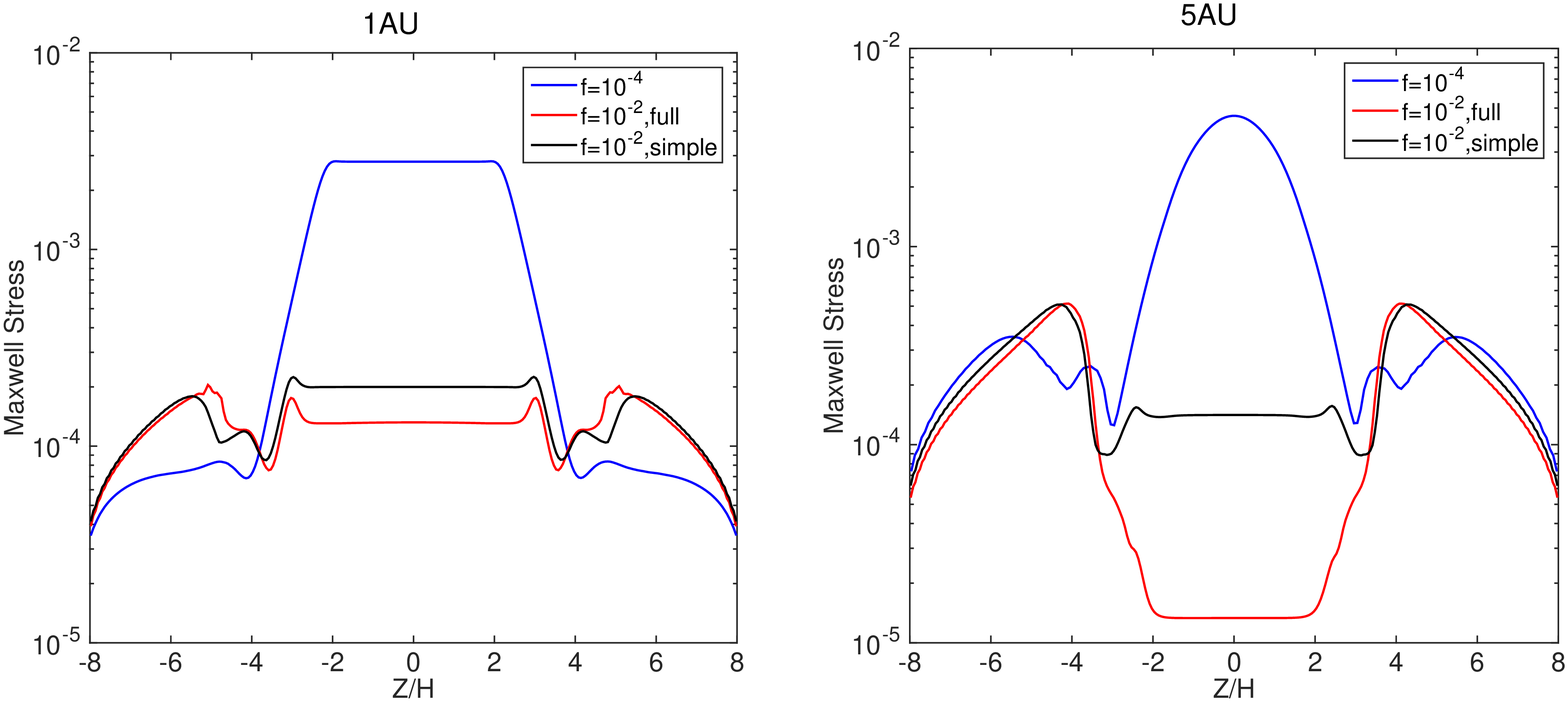}
\caption{Vertical profiles of Maxwell stress from simulations at 1 AU (left) and
5 AU (right). We have chosen grain size $a=0.1\mu$m and dust-to-gas mass
ratio $f=10^{-4}$ (blue) and $f=0.01$ (black and red), with weak net vertical
field $\beta_0=10^5$. Black lines correspond to simulations assuming
$\eta_H\propto B$ and $\eta_A \propto B^2$ in the weak field limit, while red
lines correspond to simulations using full dependence of $\eta_H$ and $\eta_A$
on $B$.}\label{fig:Maxwell}
\end{figure*}

In parallel to simulations that take into account the full dependence of
$\eta_H$, $\eta_A$ on $B$ (referred as ``full"), we also conduct simulations
assuming $\eta_H\propto B$ and $\eta_A\propto B^2$, taking the proportional
constants from the weak field limit of the full dependence (referred to as ``simple"),
and compare the difference between the two sets of runs.

Table \ref{tab:runs} lists all our quasi-1D simulations. We also show four main
simulation diagnostics at saturation: $\beta_{\rm mid}$ is the plasma $\beta$ at
disk midplane (based on total field strength), $\alpha^{\rm Max}$ is the vertically
integrated dimensionless Shakura-Sunyaev $\alpha$ based on Maxwell stress
$\int_{-z_b}^{z_b}(-B_RB_\phi)dz/\int_{-z_b}^{z_b}\rho c_s^2dz$,
$T^{z\phi}|_{z_b}$ is proportional to the torque exerted at the wind base $z_b$,
and $\dot{M}_w=\rho v_z$ is the mass loss rate measured at the top of the
simulation box (single-sided). The wind base $z_b$ is defined as the location
where rotation transitions from sub-Keplerian to super-Keplerian (in practice,
we choose it to be the first minimum of $|v_\phi-v_K|$ as one varies $z$ from
boundary towards midplane, as discussed in B14). All aspects of these diagnostic
quantities have been discussed in the literature, and this table mainly
serves as a reference.

\subsection[]{Simulation Results}

In Figure \ref{fig:simulation}, we show the simulation results with $a=0.1\mu$m,
$f=0.01$ and $\beta_0=10^5$ at 1 AU and 5 AU, respectively. The top panels
show the vertical profiles of magnetic field strength in code units, where
plasma $\beta=1$ at midplane corresponds to $B=\sqrt{2}$. As usual, toroidal
field is the dominant component, and it maintains a relatively flat profile across
the disk midplane because of excessively low ionization level there (lack of
charge carriers can not support current). We also mark the threshold field
strength $B_{\rm th}$ at which $\eta_H$ changes sign, which is computed based
on the diffusivity table directly. We have verified that the line shapes follow
exactly the corresponding curves shown in Figure \ref{fig:abundance} in
logarithmic scale. In linear scale, the minimum of $B_{\rm th}$ at $z=2-3H$
is the most prominent feature, and we see that the saturated field strength from
our simulations (dominated by toroidal field) matches closely to the minimum
value of $B_{\rm th}$. On the other hand, simulations assuming
$\eta_H\propto B$ yield stronger field around disk midplane as a result
of the HSI. This comparison clearly shows that $B_{\rm th}$ provides an
approximate upper limit that magnetic field strength can grow, resulting in
pre-matured saturation of the HSI.

The bottom panels of Figure \ref{fig:simulation} show the Elsasser number
profiles for the three non-ideal MHD effects. We first see that the locations
where $B<B_{\rm th}$ in the top panels have negative Hall Elsasser number
$\chi$ seen in the bottom panels. In other regions where $\eta_H$ is positive,
we also see that both $\eta_H$ and $\eta_A$ are reduced compared to the
``simple" case assuming $\eta_H\propto B$ and $\eta_A\propto B^2$, which
is due to a). weaker magnetic field strength and b). reduction of $|\eta_H|$
and $\eta_A$ towards stronger field.

We also performed simulations with reduced grain abundance ($a=0.1\mu$m
and $f=10^{-4}$). In this case, however, the threshold field strength is much
stronger, and we find that simulations using full diffusivity dependence on $B$
yield identical results as assuming $\eta_H\propto B$, $\eta_A\propto B^2$
scalings. The latter was adopted in \citet{Bai14}, and hence his results were
unaffected. This is because with reduced grain abundance, the threshold field
strength is much stronger (as seen in Figure \ref{fig:abundance}), and exceeds
the strength that can be attained from field amplification by the HSI.

Increasing the net vertical magnetic flux naturally leads to stronger wind, with
larger wind torque and higher mass outflow rate, as listed in Table \ref{tab:runs}.
Other than the wind being stronger, we find that the general phenomenologies
discussed above hold exactly the same way, where field amplification by the
HSI saturates at weaker field strength when full dependence of diffusivities on
$B$ is considered.

Field amplification by the HSI always weakens with
increasing grain abundance as a result of enhanced resistive dissipation.
This is already evident by comparing the effectively grain-free simulations of
\citet{Lesur_etal14} and the simulations of \citet{Bai14} with $a=0.1\mu$m and
$f=10^{-4}$, where the former showed extreme level of field amplification around
the disk midplane that approach equipartition field strength, and the latter only
found modest level of amplification. The trend continues when increasing
dust-to-gas ratio to $f=10^{-2}$, and assuming simple $\eta_H\propto B$,
$\eta_A\propto B^2$ dependence, midplane field strength is already
reduced by a factor of $\sim3$. Our simulations show that a further reduction
is achieved when full dependence of diffusivities on $B$ is considered.
This further reduction in field strength is better viewed when plotting the
Maxwell stress ($-B_RB_\phi$) profiles, which depends quadratically on field
strength. This is shown in Figure \ref{fig:Maxwell}, and the vertically integrated
values $\alpha^{\rm Max}$ are listed in Table \ref{tab:runs}.

Overall, our studies show that whether the full dependence on field strength
needs to be taken into account when small grains are abundant (or effectively,
$G\gtrsim1$ defined in (\ref{eq:g})). Reducing grain abundance leads to
larger threshold field strength $B_{\rm th}$, and and in the mean time higher
level of field amplification by the HSI. It turns out that the former increases
much faster, and the effect studied in this paper becomes essentially irrelevant
when $G\ll1$.

\section{Summary and Discussion}\label{sec:sum}

We have conducted a comprehensive study on the dependence of magnetic diffusivities
on magnetic field strength in the presence of charged grains with application to PPDs.
Ohmic resistivity $\eta_O$ is always independent of field strength. The Hall
and ambipolar diffusivities $\eta_H$ and $\eta_A$ obey simple asymptotic relations
$\eta_H\propto B$ and $\eta_A\propto B^2$ in both weak field regime where $\beta_e<1$,
and very strong field regime where $\beta_g>1$, while show complex dependence in
between (see Figure \ref{fig:analytic_diff}). In particular, when relation (\ref{eq:sign})
is satisfied (easily achieved when $n_e$ falls below $n_i$ due to the presence of
charged grains), $\eta_H$ changes sign towards strong field,
accompanied by reduction of AD. The threshold field strength $B_{\rm th}$
beyond which $\eta_H$ changes sign largely depends the ratio $n_e/n_i$, and is
weaker when $n_e/n_i$ is smaller (e.g., charged grains more dominant). Similarly, more
abundant charged grains leads to stronger reduction of $\eta_A/B^2$ between the weak
and strong field regimes.

Applying to standard MMSN model of PPDs with ionization chemistry, we find that in
order for the threshold field $B_{\rm th}$ to be well below equipartition level, a large
amount of small grains are required, as summarized in an empirical condition
$G\gtrsim1$ [defined in Equation (\ref{eq:g})], which is applicable to a wide range of
disk radii. When the same condition is satisfied, AD is also reduced substantially and
may promote the action of the MRI in the outer PPDs.

We further conducted quasi-1D shearing-box simulations for inner region of PPDs
including all three non-ideal MHD effects with net vertical field aligned with rotation.
We show that when small grains are abundant, magnetic field amplification due to
the Hall-shear instability saturates at close to the minimum value of $B_{\rm th}$.
Therefore, besides the fact that abundant small grains limit magnetic field growth by
enhancing resistivity, they also affect the gas dynamics in a more complex
way by modifying the field-strength dependence of magnetic diffusivities, especially
by enabling $\eta_H$ to change sign towards strong field.
On the other hand, with substantial grain growth (so that $G\ll1$), the threshold field
strength becomes too strong to be accessible via internal MHD processes in PPDs, and
hence it suffices to simply adopt $\eta_H \propto B$ and $\eta_A \propto B^2$ in
the weak field limit.

Our work helps clarify how the presence of grains affects ionization chemistry and
magnetic diffusivities in PPDs, and hence the general disk dynamics. This is
particularly relevant to the Hall effect: inclusion of the Hall effect not only makes
PPD gas dynamics dependent on the polarity of net vertical magnetic field, but also
leads to more sensitive dependence on grain abundance, as compared to the
Hall-free case of \citet{BaiStone13b}. In combination with other recent studies, we can
identify three regimes on the gas dynamics of inner PPDs ($\lesssim10-15$ AU)
depending on the grain abundance. 
\begin{itemize}
\item Grain-free ($G\sim0$): very strong magnetic field amplification by the HSI when
${\mb\Omega}\cdot{\mb B}>0$ \citep{Lesur_etal14,Bai14}, and bursty behavior at
$\sim5-10$ AU when ${\mb\Omega}\cdot{\mb B}<0$ \citep{Simon_etal15b}.
\item Modest grain abundance ($G<1$): modest magnetic field amplification by the HSI
when ${\mb\Omega}\cdot{\mb B}>0$, while horizontal field is reduced to close to zero
when ${\mb\Omega}\cdot{\mb B}<0$ \citep{Bai14,Bai15}.
\item High grain abundance ($G\gtrsim1$): the HSI saturates due to sign change of
$\eta_H$ at threshold field strength $B_{\rm th}$ when ${\mb\Omega}\cdot{\mb B}>0$.
\end{itemize}
Exploration of the Hall effect in PPDs is still at an early stage. There are still
pressing issues especially concerning the symmetry of disk wind, direction of
magnetic flux transport, wind kinematics, etc. The complexity introduced by grains
further adds to the richness of the overall subject of PPD gas dynamics. Conversely,
it is well known that grain size and spatial distribution in PPDs depends on
level of turbulence and global structure of the disk, and evolve with time
(e.g., \citealp{Birnstiel_etal10}). The two interrelated aspects are inherent in PPD
dynamics, and further call for more in-depth investigations in the future.

\acknowledgments

We thank K. {\"O}berg for helpful advice on disk chemistry.
XNB acknowledges support from Institute for Theory and Computation (ITC) at
Harvard-Smithsonian Center for Astrophysics.


\bibliographystyle{apj}

\label{lastpage}
\end{document}